\documentclass[12pt,a4paper]{article}

\usepackage{amsmath,amssymb}

\usepackage{geometry}
\newcommand{\klein}{\relax}

\newtheorem{lemma}{Lemma}[section]

\newtheorem{theorem}{Theorem}[section]

\newtheorem{definition}{Definition}[section]
\newtheorem{corollary}{Corollary}[section]

\let\epsilon=\varepsilon
\let\phi=\varphi

\title{Dynamics of Phase Boundary with Particle Annihilation}

\author{V.A.~Malyshev
\and A.D.~Manita\thanks{Faculty of Mathematics and Mechanics, 
Lomonosov Moscow State University,
Moscow, 119991, Russia. 
 E-mail:~malyshev2@yahoo.com, manita@mech.math.msu.su\protect\\
First published  in Proceedings of The Dobrushin International Conference - 2009, Moscow, July 2009, and in Markov Processes Related Fields, 2009, V.~15. no.~4., 575--584}}

\date{\normalsize\null\hspace*{1.64em}Date: \ \quad ~manuscript --- 2009,\newline
\null\hspace*{0.4em}ArXiv --- April 2012 \ \ \hspace*{1.08em}\null}

\begin{document}

\maketitle

We are happy to dedicate this paper to the memory of R.L.~Dobrushin.
He was a~very curious person and had wide scope in probability. Obviously
he could bring his own vision for statistical physics of economic
phenomena which is now at its very beginning.

\begin{abstract}

Infinitely many particles of two types (``plus'' and ``minus'') jump randomly along 
the one-dimensional lattice 
$\mathbf{Z}_{\varepsilon}=\varepsilon\mathbf{Z}$. Annihillations 
occur when two particles of different time occupy the same site. 
Assuming that at time $t=0$ all  ``minus'' particles are placed on 
the left of the origin and all  ``plus'' particles are  on the 
right of it, we study evolution of $\beta_\varepsilon(t)$, the 
boundary between two types. We prove that in large density limit 
$\epsilon\to 0$ the boundary $\beta_\varepsilon(t)$ converges to 
a deterministic limit. This particle system can be interpreted 
as a microscopic model of price formation on economic markets with large number of players.

\end{abstract}

{\bf Keywords: } stochastic particle systems with annihilation, scaling limits, microscopic models of price formation

{\bf MSC classes: } 60J99,
60K35,
91B26

\section{Introduction}

On one-dimensional lattice $\mathbf{Z}_{\varepsilon}=\varepsilon\mathbf{Z}=\left\{ \varepsilon m:m\in\mathbf{Z}\right\} ,\,\,\varepsilon>0,$
there are particles of two types --- ``plus particles'' and ``minus 
particles''. 
Denote by $\nu_{m}^{\pm}(t)$ the number of plus(minus)-particles at
site $\varepsilon m$ at time $t$. We define a~continuous
time Markov process on $[0,\infty)$ by the following conditions:
\begin{itemize}
\item[(1)] at time $0$ all plus particles have positive coordinates, all
minus particles have nega\-tive coordinates;
\item[(2a)] any plus particle, independently of other particles, performs
a simple random walk: that is it jumps from $\varepsilon m$ to 
$\varepsilon(m+1)$
with rate $\mu_{+}$ and from $\varepsilon m$ to $\varepsilon(m-1)$
with rate $\lambda_{+}$;
\item[(2b)] any minus particle, independently of other particles, performs
a simple random walk: that is it jumps from $\varepsilon m$ to $\varepsilon(m+1)$
with rate $\lambda_{-}$ and from $\varepsilon m$ to $\varepsilon(m-1)$
with rate $\mu_{-}$;
\item[(3a)] if a plus particle jumps to a site where there are minus particles
it immediately annihilates with one of the minus particles at this
site;
\item[(3b)] if a minus particle jumps to a site where there are plus particles
it immediately annihilates with one of the plus particles at this
site.
\end{itemize}

At any time $t>0$ the state of the process is the vector $(\nu_{m}^{\pm}(t),m\in\mathbf{Z})$.
However, it follows from (3a) and (3b) that for any $m$ and $t$ \[
\nu_{m}^{+}(t)\nu_{m}^{-}(t)=0.\]
 Moreover, all minus particles are always to the left of the leftmost
plus particle. It will be convenient to define $\beta_{\varepsilon}(t)\in\mathbf{Z}_{\varepsilon}$
as the point where the last annihilation before time $t$ happened.
We call it the phase boundary.

Note that there are no problems with the existence of this process.

Besides the interpretation related to annihilation of particles there
is another one~--- the microdynamics of the price formation, where
the market contains many players and is formed by their behaviour.
Namely, $\beta_{\varepsilon}(t)$ is the price of some product at
time $t$. Plus particles (bears) want to lower the price of this
product (we assume further on that $\alpha_{+}=\lambda_{+}-\mu_{+}>0$),
minus particles (bulls) want to increase the price (we assume $\alpha_{-}=\lambda_{-}-\mu_{-}>0$).
Annihilation is a~bargain which is performed when the demand and offer
prices meet together. Recent models of price formation \cite{Parlour,rosu_0,rosu_1,ContST}
have much in common with our model, however they are closer in spirit
to queueing models. Our model is closer to statistical physics models.
Anyway, all such models cannot pretend on practical implementation,
mainly because external influence on the action of players is not
taken into account.

We consider the large density limit $\varepsilon\to0$ under the time
scaling $t=\tau\varepsilon^{-1}$, where $t$ is microtime and $\tau$
is macrotime. Our goal is to find asymptotic behaviour of the price
$\beta(\tau)=\lim_{\varepsilon\to0}\beta_{\varepsilon}(t)$ as the
result of many micro-bargains.

\section{Main result}

\noindent\underline{{\sl Initial distribution of particles}$\vphantom{g}$}

\smallskip 
We assume for simplicity
that at time $t=0$ the distribution of plus particles is (inhomogeneous)
Poisson with density $\rho_{+}(\varepsilon m)$, where $\rho_{+}(x)$
is some strictly positive continuous function on $(0,\infty)$. This
means that the random variables $\nu_{m}^+(0)$ are independent
and have Poisson distribution with rate $\rho_{+}(\varepsilon m)$.
Similarly, the distribution of minus particles is (inhomogeneous)
Poisson with density $\rho_{-}(\varepsilon m)$, where $\rho_{-}(x)$
is some strictly positive continuous function on $(0,\infty)$.

\medskip\noindent
\underline{{\sl Notation}$\vphantom{g}$} 

\smallskip
Next we introduce main definitions and give their
intuitive interpretation. Our interpretation concerns the situation
of large density limit when, instead of point particles on a lattice,
there is a continuous media of infinitesimally small particles of
two types, the particles move with fixed velocities $-\alpha_{+}<0$, $\alpha_{-}>0$
and have initial densities $\rho_{\pm}(x)$ correspondingly. That
is there are no fluctuations. Define the functions \begin{equation}
M_{-}(r)=\int_{-r}^{0}\rho_{-}(y)\, dy\qquad\mbox{and}\qquad M_{+}(r)=\int_{0}^{r}\rho_{+}(y)\, dy\quad\mbox{for}\quad r\geq0.\label{eq:Mpm}\end{equation}
 We interprete $M_{\pm}(r)$ as the cumulative mass of plus (minus)
particles on the distance less than $r$ from zero. Under above assumptions
on $\rho_{\pm}$ we see that the functions $M_{\pm}(r)$ are strictly
increasing on $(0,+\infty)$ and, therefore, the inverse functions
$r_{\pm}(M)$, defined by the equation\[
M_{\pm}(r_{\pm}(M))=M\]
 exist and are strictly increasing. For example, the function $r_{+}=r_{+}(M)$
defines the interval $(0,r_{+})$ where the mass of plus particles
equals $M$. Then the function \begin{equation}
T(M):=\frac{r_{-}(M)+r_{+}(M)}{\alpha_{-}+\alpha_{+}}\label{eq:tM}\end{equation}
 defines the time interval $(0,T(M))$ during which mass $M$ of plus
and mass $M$ of minus particles annihilate. The function $T(M)$
is also strictly increasing on $[0,+\infty)$ and is invertible. Denote
its inverse function by $M(T)$. The place where the latter of these
particles meet\begin{equation}
r_{+}(M(T))-\alpha_{+}T=-r_{-}(M(T))+\alpha_{-}T=\beta(T)\label{eq:detm-beta}\end{equation}
 is the coordinate of the boundary at time $T$. Excluding from the
system~(\ref{eq:detm-beta}) the terms that are linear in $T$, we
get \[
\beta(T)=r_{+}(M(T))\frac{\alpha_{-}}{\alpha_{-}+\alpha_{+}}-r_{-}(M(T))\frac{\alpha_{+}}{\alpha_{-}+\alpha_{+}}\,.\]

\medskip\noindent
\underline{{\sl Scaling limit for the stochastic model}$\vphantom{g}$}

\smallskip 
Here we return to the
stochastic particle model and formulate the main result.

\begin{theorem} For any fixed $\tau\geq0$ the following convergence
in probability holds\[
\beta_{\varepsilon}(\varepsilon^{-1}\tau)\rightarrow\beta(\tau)\qquad\qquad(\varepsilon\rightarrow0),\]
 where the function $\beta:\,\,\mathbf{R}_{+}\rightarrow\mathbf{R}$
is deterministic and has the following explicit form\[
\beta(\tau)=\frac{-\alpha_{+}r_{-}(M(\tau))+\alpha_{-}r_{+}(M(\tau))}{\alpha_{-}+\alpha_{+}}\,.\]
 \end{theorem}

 \begin{corollary}
Consider the homogeneous case $\rho_{-}(y)\equiv\rho_{-},$ $y<0,$
$\rho_{+}(y)\equiv\rho_{+},$ $y>0.$
All functions defined above are linear: $M_{-}(r)=\rho_{-}r,$ 
$M_{+}(r)=\rho_{+}r,$ $r_{\pm}(M)=M/\rho_{\pm},$ 
\begin{align*}
T(M)&=M\,\frac{\rho_{-}^{-1}+\rho_{+}^{-1}}{\alpha_{-}+\alpha_{+}},\\
M(T)&=T\frac{\alpha_{-}+\alpha_{+}\vphantom{A^B}}%
{\rho_{-}^{-1}+\rho_{+}^{-1}}\,,
\end{align*}
and, hence, the phase boundary $\beta_{\varepsilon}(\tau\varepsilon^{-1})$
moves with an asymptotically constant velocity:
\begin{align*}
\beta_{\varepsilon}(\tau\varepsilon^{-1})\rightarrow\beta(\tau)&=
\tau\,\frac{-\alpha_{+}\rho_{-}^{-1}+\alpha_{-}\rho_{+}^{-1}}{\rho_{-}^{-1}
+\rho_{+}^{-1}}\\
&=\tau\,\frac{-\alpha_{+}\rho_{+}+\alpha_{-}\rho_{-}}{\rho_{+}+\rho_{-}}\,.
\end{align*}
\end{corollary}

\section{Proof}

Our plan is to show that the limiting behavior of $\beta_{\varepsilon}(t)$
in the stochastic model corresponds to the deterministic evolution
described in (\ref{eq:Mpm})--(\ref{eq:detm-beta}). To do this we
need some control over the random fluctuations in the limit $\varepsilon\rightarrow0$.
Now we fix $M$ and consider the following random variables 
$A_{i,\pm}=A_{i,\pm}(M),\, i=0,1,2$, 
(we will prove that they are of the order $o(\varepsilon^{-1})$):

\begin{enumerate}
\item Denote by $Q_{\varepsilon,\pm}=Q_{\varepsilon,\pm}(M)$ the number of
$(\pm)$-particles which were at time $t=0$ correspondingly in the
intervals
\begin{equation}
I_{+}^{\circ}=\left(0,r_{+}(M)\right)\cap\mathbf{Z}_{\varepsilon}\quad
\textrm{and}\quad 
I_{-}^{\circ}=\left(-r_{-}(M),0\right)\cap\mathbf{Z}_{\varepsilon} .
\label{eq:I0-I0+}
\end{equation}
 Define $A_{0,\pm}=Q_{\varepsilon,\pm}-M\varepsilon^{-1}$. 
\item All particles among them will be annihilated
during time $t=T(M)\varepsilon^{-1}$ except of the number $A_{1,\pm}$
of them. 
\item Define $A_{2,\pm}$ as the number of plus and minus particles which were not
at time $t$ in the intervals (\ref{eq:I0-I0+}) but were annihilated
during time $t=T(M)\varepsilon^{-1}$. 
\end{enumerate}

This control can be achieved by use of exponential bounds for some
families of events. The proof uses some ideas from \cite{MalZam}.

\begin{definition} We say that a family of events $\mathcal{A}=\left\{ A_{\varepsilon}\right\} _{\varepsilon>0}$
has a property of exponential asymptotic sureness (e.a.s.) if there
exist constants $\mathcal{K}_{\mathcal{A}}>0$, $q_{\mathcal{A}}>0$,
$\varepsilon_{\mathcal{A}}>0$ such that for all $\varepsilon<\varepsilon_{\mathcal{A}}$
the following inequality holds 
\[
\mathsf{P}\left(A_{\varepsilon}\right)\geq1-\mathcal{K}_{\mathcal{A}}
\exp\left(-q_{\mathcal{A}}\varepsilon^{-1}\right).
\]
 \end{definition}

In the sequel, for breavity, we say sometimes that the event $A_{\varepsilon}$
has probability exponentially close to one. We will use the following
fact: if two sequences $\mathcal{A}=\left\{ A_{\varepsilon}\right\} _{\varepsilon>0}$
and $\mathcal{B}=\left\{ B_{\varepsilon}\right\} _{\varepsilon>0}$
have the property e.a.s., then this property holds also for the sequence
$\mathcal{C}=\left\{ A_{\varepsilon}\cap B_{\varepsilon}\right\} _{\varepsilon>0}$.

It is helpful to enumerate the particles at time $0$ somehow with
the only condition that \[
\cdots\leq x_{3}^{-}(0)\leq x_{2}^{-}(0)\leq x_{1}^{-}(0)<0<x_{1}^{+}(0)\leq x_{2}^{+}(0)\leq x_{3}^{+}(0)\leq\cdots\qquad.\]
 Denote by $q_{-}(1)$ and $q_{+}(1)$ the indices of plus and minus
particles of the first annihilating pair. One can assume that if some
plus (minus) particle jumps to a site where there are several minus
(plus) particles then it annihilates with the minus (plus) particle
having minimal index. Let $\sigma_{1}$ be the time moment when the
first annihilation occurs. Since particles move independently, their
order can change in time, so, in general, $x_{q_{-}(1)}^{-}(0)\not=x_{1}^{-}(0)$
and $x_{q_{+}(1)}^{+}(0)\not=x_{1}^{+}(0)$. Similarly, we define
$q_{-}(m)$ and $q_{+}(m)$ as the indices of the particles of the
$m$-th annihilating pair and $\sigma_{m}$ as the time moment of
the $m$-th annihilation.

Fix some $M>0$. Let $N_{\varepsilon}=[M\varepsilon^{-1}]$. Consider
the $N_{\varepsilon}$-th pair of annihilating particles, $x_{q_{-}(N_{\varepsilon})}^{-}$
and $x_{q_{+}(N_{\varepsilon})}^{+}$. The main idea is to prove that
for small $\varepsilon$ the random time $\sigma_{N_{\varepsilon}}$
is close to the value $T(M)\varepsilon^{-1}$ and the random coordinate
$x_{q_{-}(N_{\varepsilon})}^{-}(0)\in\mathbf{Z}_{\varepsilon}$ is
close to $-r_{-}(M)$. In more precise terms, it is sufficient to
prove that for any small fixed positive numbers $\varkappa_{0},\varkappa_{1},\zeta_{-},\zeta_{+}$
\emph{with probability exponentially close to one} (as $\varepsilon\rightarrow0$)
the following holds:

\begin{itemize}
\item[(a)] the moment $\sigma_{N_{\varepsilon}}$ belongs to the time
interval $(t_{0}(M,\varepsilon),t_{1}(M,\varepsilon))$, where 
\begin{align}
t_{0}(M,\varepsilon)&= (T(M)-\varkappa_{0})\varepsilon^{-1},\notag \\ 
t_{1}(M,\varepsilon)&=(T(M)+\varkappa_{1})\varepsilon^{-1};\label{eq:t0-re}
\end{align}
\item[(b)] the starting point of the minus particle $x_{q_{-}(N_{\varepsilon})}^{-}(0)$
belongs to the set $(-r_{-}(M)-\zeta_{-},-r_{-}(M)+\zeta_{-})\cap\mathbf{Z}_{\varepsilon}$;
\item[(c)] similarly, the starting point of the plus particle $x_{q_{+}(N_{\varepsilon})}^{+}(0)$
belongs to the set $(r_{+}(M)-\zeta_{+},r_{+}(M)+\zeta_{+})\cap\mathbf{Z}_{\varepsilon}$.
\end{itemize}

Let us prove the theorem assuming that the above statements (a)--(c)
are proved. Recall that $\beta_{\varepsilon}(\sigma_{N_{\varepsilon}}+0)=x_{q_{-}(N_{\varepsilon})}^{-}(\sigma_{N_{\varepsilon}})=x_{q_{+}(N_{\varepsilon})}^{+}(\sigma_{N_{\varepsilon}})$.
Individual motion of a minus particle is a simple random walk on $\mathbf{Z}_{\varepsilon}$
with the mean drift $\alpha_{-}\varepsilon=\left(\lambda_{-}-\mu_{-}\right)\varepsilon$,
so applying the upper bound of the large deviation theory, we get
that for any fixed $i\in\mathbf{N}$, $s>0$ and $\delta_{0}>0$ with
probability exponentially close to one \[
x_{i}^{-}(s\varepsilon^{-1})-x_{i}^{-}(0)\in\left(\left(\alpha_{-}-\delta_{0}\right)s,\left(\alpha_{-}+\delta_{0}\right)s\right).\]
 In fact, even stronger result holds: for fixed $s_{2}>s_{1}>0$ and
$\delta_{0}>0$ the family of events $\left\{ D_{\varepsilon}\right\} $,
where 
$$
D_{\varepsilon}=\left\{ \, x_{i}^{-}(s\varepsilon^{-1})-x_{i}^{-}(0)\in
\left(\left(\alpha_{-}-\delta_{0}\right)s,\left(\alpha_{-}+\delta_{0}\right)
s\right),\; \forall s\in[s_{1},s_{2}]\,\right\} ,
$$
has a property of e.a.s. Together with (a) this gives 
\[
x_{q_{-}(N_{\varepsilon})}^{-}(\sigma_{N_{\varepsilon}})-x_{q_{-}(N_{\varepsilon})}^{-}(0)\in\left(\left(\alpha_{-}-\delta_{0}\right)\left(T(M)-\varkappa_{0}\right),\left(\alpha_{-}+\delta_{0}\right)\left(T(M)+\varkappa_{1}\right)\right)\]
 with probability exponentially close to one. Combining the latter
statement with the statement (b) we conclude that with probability
exponentially close to one 
\[
x_{q_{-}(N_{\varepsilon})}^{-}(\sigma_{N_{\varepsilon}})\in
\left(\alpha_{-}T(M)-r_{-}(M)-\gamma,\alpha_{-}T(M)-r_{-}(M)+\gamma\right)
\]
 where $\gamma=\gamma(\delta_{0},\varkappa_{0},\varkappa_{1},\zeta_{-})>0$
can be made arbitrary small, i.e., 
$$\gamma(\delta_{0},\varkappa_{0},\varkappa_{1},\zeta_{-})\rightarrow0 \quad
\mbox{ as } \max(\delta_{0},\varkappa_{0},\varkappa_{1},\zeta_{-})\rightarrow0.$$
Using (\ref{eq:tM}) we see that 
\begin{align*}
-r_{-}(M)+\alpha_{-}T(M) & =  -r_{-}(M)+\alpha_{-}\frac{r_{-}(M)+r_{+}(M)}{\alpha_{-}+\alpha_{+}}\\
 & =  -r_{-}(M)\frac{\alpha_{+}}{\alpha_{-}+\alpha_{+}}+r_{+}(M)\frac{\alpha_{-}}{\alpha_{-}+\alpha_{+}}\,
\end{align*}
 and, hence, 
$$
\beta_{\varepsilon}(\sigma_{N_{\varepsilon}})\rightarrow-r_{-}(M)
\frac{\alpha_{+}}{\alpha_{-}+\alpha_{+}}+r_{+}(M)
\frac{\alpha_{-}}{\alpha_{-}+\alpha_{+}}\,
$$
in probability as $\varepsilon\rightarrow0$.

To finish the proof of theorem we need only to check that 
$$
\beta_{\varepsilon}(\sigma_{N_{\varepsilon}})-\beta_{\varepsilon}
(\varepsilon^{-1}T(M))\rightarrow 0
\quad
\mbox{ as } \varepsilon\rightarrow 0.
$$ 
This corresponds to continuity property
of the border on the macroscopic time scale~$\tau=T$. To establish
this fact we should take into account that: 1) due to the drift assumption
($\alpha_{\pm}>0$) the random sequence $\left\{ \sigma_{m+1}-\sigma_{m},\,\, m\in\mathbf{N}\right\} $
admits uniform exponential estimates for the tails of the distribution
functions of $\sigma_{m+1}-\sigma_{m}$ (we refer the reader to~\cite{1995_book_02}
for the corresponding techniques); 2) in finite microtime~$t$ the
displacements of walking particles have the order $O(\varepsilon)$
while in finite macrotime~$\tau$ their displacements have the order
$O(1)$. We omit the details.

To prove the statements (a) we need the following main lemma.
Denote by $\mathcal{N}_{-}(0,t_{m}(M,\varepsilon))$ a set of minus
particles that collide with plus particles on the time interval $(0,t_{m}(M,\varepsilon))$.

\begin{lemma}\label{l-N-t0t1}For any sufficiently small $\varkappa_{2},\varkappa_{3}>0$
the following events 
\begin{align*}
F_{\varepsilon}&=\left\{\left|\mathcal{N}_{-}(0,t_{0}(M,\varepsilon))\right|
<(M-\varkappa_{2})\varepsilon^{-1}\right\},\\
G_{\varepsilon}&=\left\{\left|\mathcal{N}_{-}(0,t_{1}
(M,\varepsilon))\right|>(M+\varkappa_{3})\varepsilon^{-1}\right\}
\end{align*}
 have the probabilities exponentially close to one.
\end{lemma}

Lemma~\ref{l-N-t0t1} follows from Lemmas~\ref{l-ini-distr} and 
\ref{l-all-col}. Lemma~\ref{l-ini-distr}
deals with the initial distribution of particles and Lemma~\ref{l-all-col}
controls the deplacements of minus and plus particles.

\begin{lemma}\label{l-ini-distr}Let $y_{1}<y_{2}\leq0$ and $0\leq z_{1}<z_{2}$.
Then for any $\delta>0$ the following families of events
have probabilities exponentially close to one: 
\begin{align*}
L_{\varepsilon}&= \{ \mbox{the number of minus particles sitting at time
$t=0$ in the set }  \\
&\quad\;\;\left(y_{1},y_{2}\right)\cap\mathbf{Z_{\varepsilon}} 
\mbox{ is between }
\textstyle{\bigl(\int_{y_{1}}^{y_{2}}\rho_{-}(y)\, dy
-\delta\bigr)\varepsilon^{-1}\,} \\
& \quad\;\, \mbox{ and } \textstyle{\bigl(\int_{y_{1}}^{y_{2}}\rho_{-}(y)\, dy
+\delta\bigr)
\varepsilon^{-1} \} ,} \\
R_{\varepsilon}&= \{ \mbox{the number of plus particles sitting at time
$t=0$ in the set } \\ 
&\quad\;\;\left(z_{1},z_{2}\right)\cap\mathbf{Z_{\varepsilon}} \mbox{ is between }
\textstyle{\bigl(\int_{z_{1}}^{z_{2}}\rho_{+}(y)\, dy-\delta\bigr)
\varepsilon^{-1}\,} \\
&\quad\;\, \mbox{ and } \textstyle{\bigl(\int_{z_{1}}^{z_{2}}\rho_{+}(y)\, dy+
\delta\bigr)
\varepsilon^{-1}\}} . 
\end{align*}
\end{lemma}

\begin{lemma}\label{l-all-col}For any $\delta_{1}>0$ each family
of events
\begin{align*}
A_{\varepsilon}&=\{ \mbox{all particles } 
x_{k}^{\pm}(0)\in\left(-r_{-}(M)+\delta_{1}\,,\, r_{+}(M)-\delta_{1}\right)\cap\mathbf{Z_{\varepsilon}}
\mbox { collide} \\
 &\quad\; \mbox{ with particles of opposite sign till the time moment }
t(M)\varepsilon^{-1}\}, \\
B_{\varepsilon}&=\{ \mbox{on the time interval } 
t\in(0,s\varepsilon^{-1})
\mbox{ none of minus particles, started at }  \\
&\quad\;\;\,t=0 \mbox{ from the set }
(-\infty,-r_{-}(M)-\delta_{1})\cap\mathbf{Z_{\varepsilon}},
\mbox{ collides with any } \\
&\quad\;\mbox{ plus particle, started at $t=0$ from the set } 
(r_{+}(M)+\delta_{1},+\infty)\cap\mathbf{Z_{\varepsilon}} \} , 
\end{align*}
satisfies the e.a.s.\ property. Moreover, fix any $y,\varkappa>0$
and consider the following subsets of $\mathbf{Z_{\varepsilon}}$:
\[
S_{2}^{-}=\left(-\infty,-y-\varkappa\right),
\quad S_{1}^{-}=\left(-y,0\right),
\quad S_{1}^{+}=\left(0,y\right),\quad 
S_{2}^{+}=\left(y+\varkappa,+\infty\right).
\] 
Define the events 
\begin{align*}
V_{\varepsilon}&=\{ 
\mbox{on the time interval 
$t\in(0,s\varepsilon^{-1})$
 none of minus particles, started at }\\
& \quad\;\; t=0\, \mbox{ from } S_{2}^{-},\! \mbox { will
 meet some minus particle, started from the set } S_{1}^{-} \} 
; \\
U_{\varepsilon}&= \{ 
\mbox{on the time interval $t\in(0,s\varepsilon^{-1})$
none of plus particles, started at }\\
&\quad\;\; t=0\, \mbox{ from $S_{2}^{+}$, will
meet some minus particle, started from the set $S_{1}^{+}$}\}. 
\end{align*}
 Then the families of events $\left\{ V_{\varepsilon}\right\} $ 
and $\left\{ U_{\varepsilon}\right\} $
 have the e.a.s.\ property.
\end{lemma}

Proofs of Lemmas~\ref{l-ini-distr} and~\ref{l-all-col}
are based on standard probabilistic methods~\cite{1995_book_02}
and are omitted. Let us explain now how using these two lemmas one
can get, for example, the upper bound for $\left|\mathcal{N}_{-}(0,t_{0}(M,\varepsilon))\right|$
in Lemma~\ref{l-N-t0t1}.

Firstly, we include in this bound all minus particles starting
at $t=0$ from the set $\left(-r_{-}\left(\, M(T(M)-\varkappa_{0})\,\right)-\delta_{5},0\right)$
where $\delta_{5}>0$ is small and will be fixed later. By Lemma~\ref{l-ini-distr}
there is no more than $\left(M_{-}\left(r_{-}\left(\, M(T(M)-\varkappa_{0})\,\right)+\delta_{5}\right)+\delta_{6}\right)\varepsilon^{-1}$
of such particles e.a.s.\ for small $\delta_{6}>0$.

We should add to this bound all minus particles that started
at $t=0$ from the set $\left(-\infty,-r_{-}\left(\, M(T(M)-\varkappa_{0})\,\right)-\delta_{5}\right)$
and annihilated in the time interval $(0,t_{0}(M,\varepsilon))$ with
some plus particles. We will show now that with probability exponentially
close to one the number $N^{\circ}(0,t_{0}(M,\varepsilon))$ of such
minus particles can be estimated as $c\varepsilon^{-1}$ where $c>0$
is any prefixed small constant. Indeed, by Lemma~\ref{l-all-col}
(again in the sense of e.a.s.) the minus particles in question can
annihilate only by colliding with some plus particles, started at
$t=0$ from the set $\left(0,r_{+}\left(\, M(T(M)-\varkappa_{0})\,\right)+\delta_{5}\right)$.
By Lemma~\ref{l-ini-distr}, the number of the plus particles in this
set is bounded by $\left(M_{+}\left(r_{+}\left(\, M(T(M)-\varkappa_{0})\,\right)+\delta_{5}\right)+\delta_{6}\right)\varepsilon^{-1}$.
From this bound we should exclude plus particles which was annihilated
in collisions with minus particles started from the set $\left(-r_{-}\left(\, M(T(M)-\varkappa_{0})\,\right)+\delta_{5},0\right)$,
since by the part~{}``$V_{\varepsilon}$'' of Lemma~\ref{l-all-col}
during the time interval $(0,t_{0}(M,\varepsilon))$ the latter minus
particles will go ahead of the minus particles started from $\left(-\infty,-r_{-}\left(\, M(T(M)-\varkappa_{0})\,\right)-\delta_{5}\right)$.
By Lemma~\ref{l-ini-distr}, initially there was no less than 
$\left(M_{-}\left(r_{-}\left(\, M(T(M)-\varkappa_{0})\,\right)-
\delta_{5}\right)-\delta_{6}\right)\varepsilon^{-1}$
particles
 in the set 
$$\left(-r_{-}\left(\, M(T(M)-\varkappa_{0})\,\right)+\delta_{5},0\right).
$$ 
So using the mean value theorem from analysis we get 
\begin{align*}
\varepsilon\cdot N^{\circ}(0,t_{0}(M,\varepsilon)) & \leq 
 \left(M_{+}\left(r_{+}\left(\, M(T(M)-\varkappa_{0})\,\right)+
\delta_{5}\right)+\delta_{6}\right)\\
 &\quad {} -\left(M_{-}\left(r_{-}\left(\, M(T(M)-\varkappa_{0})\,\right)
-\delta_{5}\right)-\delta_{6}\right)\\
 & = M(T(M)-\varkappa_{0})+M_{+}'(\theta_{1})\delta_{5}+\delta_{6}\\
&\quad {} -\left(\left(\, M(T(M)-\varkappa_{0})\,\right)-
M_{-}'(\theta_{2})\delta_{5}-\delta_{6}\right)\\
 & \leq \delta_{5}(\|M_{-}'\|_{C}+\|M_{+}'\|_{C})+2\delta_{6}.
\end{align*}
 Hence, in the sense of e.a.s., 
\begin{align*}
\varepsilon\cdot\left|\mathcal{N}_{-}(0,t_{0}(M,\varepsilon))\right| & \leq 
 (M_{-}(r_{-}(M(T(M)-\varkappa_{0}))\\
&\quad {} +\delta_{5})+\delta_{6})+
\delta_{5}(\|M_{-}'\|_{C}+\|M_{+}'\|_{C})+2\delta_{6} \\
 & =  M(T(M)-\varkappa_{0})+M_{-}'(\theta_{3})\delta_{5}\\
&\quad {} +
\delta_{5}(\|M_{-}'\|_{C}+\|M_{+}'\|_{C})+3\delta_{6}\\
 & \leq M(T(M)-\varkappa_{0})+\delta_{5}(2\|M_{-}'\|_{C}+\|M_{+}'\|_{C})+
3\delta_{6} .
\end{align*}
 It follows from~(\ref{eq:tM}) and assumptions on $\rho_{\pm}$
that $M'(t)\geq k$ for some $k>0$. Therefore, $M(T(M)-\varkappa_{0})\leq M-k\varkappa_{0}$.
Given $\varkappa_{0}>0$ we are allowed to chose positive constants
$\delta_{5}$ and $\delta_{6}$ as small as we like. So, finally,
we get that with probability exponentially close to one the following
estimate holds 
\[
\left|\mathcal{N}_{-}(0,t_{0}(M,\varepsilon))
\right|\leq\Bigl(M-\frac{k\varkappa_{0}}{2}\Bigr)\varepsilon^{-1}\,.
\]
 Lower bound for $\left|\mathcal{N}_{-}(0,t_{1}(M,\varepsilon))\right|$
can be obtained in a similar way.

 To get the proof of statement (b) one should combine
Lemma~\ref{l-all-col} with the following lemma.

\begin{lemma}For any $\varkappa_{5}>0$, the event 
\[
H_{\varepsilon}= \left\{ x_{i}^{-}(0)\in\left(-(1+\varkappa_{5})r_{-}(M),
0\right)\quad\forall i\in\mathcal{N}_{-}(0,t_{1}(M,\varepsilon))\right\}
\]
 has the property of e.a.s.
\end{lemma} 

Proof of this lemma uses arguments similar to the proof
of Lemma~\ref{l-N-t0t1}. The statement (c) is just a symmetric
modification of the statement (b).

\end{document}